\documentclass{emulateapj}
\usepackage{lscape}
\usepackage{times}
\usepackage{graphicx}% Include figure files

\shorttitle{Fermi-LAT detection of HESS~J1825$-$137}
\shortauthors{Grondin et al.}

\begin{document}

\title{Detection of the Pulsar Wind Nebula HESS~J1825$-$137\\
	with the \emph{Fermi} Large Area Telescope }


\author{
M.-H.~Grondin\altaffilmark{2,1,*}, 
S.~Funk\altaffilmark{3,1}, 
M.~Lemoine-Goumard\altaffilmark{2,1}, 
A.~Van~Etten\altaffilmark{3,1}, 
J.~A.~Hinton\altaffilmark{4}, 
F.~Camilo\altaffilmark{5}, 
I.~Cognard\altaffilmark{6}, 
C.~M.~Espinoza\altaffilmark{7}, 
P.~C.~C.~Freire\altaffilmark{8}, 
J.~E.~Grove\altaffilmark{9}, 
L.~Guillemot\altaffilmark{8}, 
S.~Johnston\altaffilmark{10}, 
M.~Kramer\altaffilmark{7,8}, 
J.~Lande\altaffilmark{3}, 
P.~Michelson\altaffilmark{3}, 
A.~Possenti\altaffilmark{11}, 
R.~W.~Romani\altaffilmark{3}, 
J.~L.~Skilton\altaffilmark{12}, 
G.~Theureau\altaffilmark{6}, 
P.~Weltevrede\altaffilmark{7}
}
\altaffiltext{1}{Corresponding authors: S.~Funk, funk@slac.stanford.edu; M.-H.~Grondin, grondin@cenbg.in2p3.fr or grondin@astro.uni-tuebingen.de; M.~Lemoine-Goumard, lemoine@cenbg.in2p3.fr; A.~Van Etten, ave@stanford.edu.}
\altaffiltext{2}{Universit\'e Bordeaux 1, CNRS/IN2P3, Centre d'\'Etudes Nucl\'eaires de Bordeaux Gradignan, 33175 Gradignan, France}
\altaffiltext{3}{W. W. Hansen Experimental Physics Laboratory, Kavli Institute for Particle Astrophysics and Cosmology, Department of Physics and SLAC National Accelerator Laboratory, Stanford University, Stanford, CA 94305, USA}
\altaffiltext{4}{Department of Physics and Astronomy, University of Leicester, Leicester, LE1 7RH, UK}
\altaffiltext{5}{Columbia Astrophysics Laboratory, Columbia University, New York, NY 10027, USA}
\altaffiltext{6}{Laboratoire de Physique et Chimie de l'Environnement, LPCE UMR 6115 CNRS, F-45071 Orl\'eans Cedex 02, and Station de radioastronomie de Nan\c{c}ay, Observatoire de Paris, CNRS/INSU, F-18330 Nan\c{c}ay, France}
\altaffiltext{7}{Jodrell Bank Centre for Astrophysics, School of Physics and Astronomy, The University of Manchester, M13 9PL, UK}
\altaffiltext{8}{Max-Planck-Institut f\"ur Radioastronomie, Auf dem H\"ugel 69, 53121 Bonn, Germany}
\altaffiltext{9}{Space Science Division, Naval Research Laboratory, Washington, DC 20375, USA}
\altaffiltext{10}{Australia Telescope National Facility, CSIRO, Epping NSW 1710, Australia}
\altaffiltext{11}{INAF - Cagliari Astronomical Observatory, I-09012 Capoterra (CA), Italy}
\altaffiltext{12}{School of Physics and Astronomy, University of Leeds, Leeds LS2 9JT, UK}
\altaffiltext{*}{Now at : Institut f\"ur Astronomie und Astrophysik T\"ubingen, Universit\"at T\"ubingen, Sand 1, D-72076 T\"ubingen, Germany}

\begin{abstract}
  We announce the discovery of 1~--~100 GeV gamma-ray emission from the archetypal 
TeV pulsar wind nebula HESS~J1825$-$137 using 20~months of survey data from the {\it Fermi}
  Large Area Telescope (LAT). The gamma-ray emission detected by the
  LAT is significantly spatially extended, with a best-fit rms
  extension of $\sigma = 0.56^{\circ} \pm 0.07^{\circ}$ for an assumed
  Gaussian model. The 1 GeV~--~100~GeV LAT spectrum of this source is
  well described by a power-law with a spectral index of $1.38 \pm
  0.12 \pm 0.16$ and an integral flux above 1 GeV of $(6.50 \pm 0.21 \pm
  3.90) \times 10^{-9}$ cm$^{-2}$~s$^{-1}$. The first errors represent
  the statistical errors on the fit parameters, while the second ones
  are the systematic uncertainties. Detailed morphological and
  spectral analyses bring new constraints on the energetics and
  magnetic field of the pulsar wind nebula system. The spatial extent and 
hard spectrum of the GeV emission are consistent with the picture of an inverse Compton 
origin of the GeV-TeV emission in a cooling-limited nebula powered by the pulsar PSR~J1826$-$1334.
\end{abstract}

\keywords{pulsar, pulsar wind nebula, ISM: individual objects
  (G18.0-0.7, HESS~J1825$-$137), pulsars: individual (PSR B1823$-$13, PSR
  J1826$-$1334)}

\section{Introduction}

Pulsars dissipate their energy via magnetized particle winds
(consisting of electron/positron pairs) with the confinement of this
particle wind-outflow leading to the phenomenon of pulsar wind nebulae
(PWNe). PWNe develop when the particle wind collides with
its surroundings, in particular with the slowly expanding supernova
ejecta, and form a termination shock. Even though many PWNe are 
bright enough to be resolvable both spatially and spectrally
at wavelengths from radio to very high
energy (VHE; $E > 100$~GeV) gamma-rays, many open questions remain. How is the
rotational energy of the pulsar converted into a relativistic particle
wind? What is the bulk Lorentz factor of the wind, what is the
mechanism by which the particles are accelerated at the termination
shock? What is the partition between magnetic and particle energy in
the wind? To address these questions, PWNe can be seen as
laboratories for relativistic astrophysics with the advantage of
having a well-localized (and usually well-characterized) central
energy-source, the spin-down power of a neutron star.

Many studies of PWNe have been conducted in the X-ray
band with instruments such as {\emph{XMM-Newton}} or {\emph{Chandra}}~\citep{kargaltsev}, 
but recent gamma-ray observations with H.E.S.S., MAGIC and VERITAS have
contributed significantly to the understanding of PWNe. Together with the assumption
that the X-ray emission is synchrotron whereas the higher energies are
Inverse Compton (IC) emission, these observations allow us to derive the magnetic field 
strength in the nebula. For the PWNe seen in TeV gamma-rays,
typical magnetic field strengths are $\sim 10$~$\mu$G~\citep{okkie}.
In turn, the lifetime of the synchrotron-emitting electrons can be
determined from the magnetic field strengths. For keV X-ray emitting electrons 
the lifetime of less than 100 years is significantly shorter than
the age of the central pulsar ($10^3-10^5$ years) whereas the TeV 
emitting electrons have lifetimes of order 10$^4$ years, comparable to the 
pulsar age. This has the consequence that the X-ray PWNe should be
spatially very compact, whereas the PWNe as seen in TeV should be 
significantly larger.
The discovery of energy-dependent morphology
at TeV energies in the archetypal system HESS~J1825$-$137 (surrounding
the energetic pulsar PSR\,J1826$-$1334) confirmed this basic picture
by demonstrating that the emission is dominated by `relic' electrons
from the earlier epochs of the nebula in which the pulsar was spinning
down more rapidly, releasing more energy into the system~\citep{HESSJ1825}. The
gamma-ray studies of PWNe also showed that the overall efficiency of
conversion of rotational energy into TeV emission in these systems is
extremely high ($\gg10\%$ in some cases).

The recently-launched Large Area Telescope (LAT) on board the
{\emph{Fermi}} Gamma-ray Space Telescope has established
pulsars as the most numerous class of identified Galactic GeV
gamma-ray emitting objects~\citep{FirstCat} and has also shown that up to 10\%
of the rotational power from pulsars is emitted in the Fermi-LAT
energy band. The spectrum of the pulsed component cuts off in the
sub-10-GeV range~\citep{psrcat}. As detailed above, the spectra of PWNe extend out
to TeV energies making PWNe the most numerous Galactic objects
at those energies.
While for a few TeV sources there is a clear
morphological match (e.g.\ with the X-ray emission) that leads to the
identification with a PWN, for most TeV sources the situation is less
obvious. Generally, a positional match with a pulsar (along with the
requirement that the energetics of the pulsar is sufficient to power
the TeV emission) is used to suggest an association. The situation
is, however, further complicated by the fact that for many of the
TeV-detected PWNe, in particular for the ones that are associated with
middle-aged rather than young pulsars, the gamma-ray emission is offset
from the pulsar.

HESS~J1825$-$137 is one of the prime examples for such an offset PWN
system. The system seems to be powered by the energetic radio pulsar
PSR\,J1826$-$1334. High resolution X-ray
observations of the PWN with {\emph{XMM-Newton}} showed a compact core with a
hard photon index ($\Gamma = 1.6^{+0.1}_{-0.2}$) of size 30$''$ embedded
in a larger diffuse structure of extension $\sim 5'$ extending to the
south of the pulsar with a softer photon index of $\Gamma \sim
2.3$~\citep{gaensler03}. The TeV gamma-ray emission has a much larger extent ($\sim
0.5^{\circ}$) and shows a similar softening of the photon index from
$\Gamma=2.0$ close to the pulsar to $\Gamma=2.5$ at a distance of
$1^{\circ}$ to the south of the pulsar~\citep{HESSJ1825}. Based on hydrodynamical
simulations~\citep{Blondin} the asymmetric nature of the emission can
be explained by dense interstellar material to the north of the
pulsar. The reverse shock of the supernova explosion from that
direction might have been pushed inward, interacting therefore
relatively early with the PWN and thus pushing the X-ray and TeV
gamma-ray emission mainly to the south.

Here we report on GeV gamma-ray observations of the
HESS~J1825$-$137/PSR~J1826$-$1334 system with \emph{Fermi}-LAT. 

%%%%%%%%%%%%%%%%%%%%%%%%%%%%%%%%%%%%%%%%%%%%%%%%%%%%%%%%%%%%%%%%%%%%%%%%%%%%%%%%%%%%%%%%%%%%%%%%%%%%%%%%%%%%%
\section{LAT description and observations}
\label{lat}
The LAT is a gamma-ray telescope that detects photons by conversion
into electron-positron pairs and operates in the energy range between
20 MeV and more than 300 GeV. It is made of a high-resolution converter tracker
(for direction measurement of the incident gamma-rays), a CsI(Tl) crystal
calorimeter (for energy measurement) and an anti-coincidence detector to
identify the background of charged particles \citep{Atwood et
  al. 2009}. In comparison to EGRET, the LAT has a larger effective
area ($\sim$ 8000 cm$^{2}$ on-axis above 1~GeV), a broader field of
view ($\sim$ 2.4 sr) and a superior angular resolution ($\sim$
0.6$^{\circ}$ 68$\%$ containment at 1 GeV for events converting in the
front section of the tracker). Details of the instrument and data
processing are given in~\citet{Atwood et al. 2009}. The on-orbit
calibration is described in \cite{VelaPulsarLAT}.

The following analysis was performed using 20 months of data collected
starting August 4, 2008, and extending until April 21, 2010. Only
gamma-rays in the Pass~6 {\emph{Diffuse}} class events were selected (i.e. with 
the tightest background rejection), and from this sample, we excluded 
those coming from a zenith angle larger than 105$^{\circ}$ to the 
detector axis because of the possible contamination from secondary gamma-rays 
from the Earth's atmosphere~\citep{FermiAlbedo}. We have used P6$\_$V3 post-launch 
instrument response functions (IRFs), which take into account pile-up
and accidental coincidence effects in the detector
subsystems.

%%%%%%%%%%%%%%%%%%%%%%%%%%%%%%%%%%%%%%%%%%%%%%%%%%%%%%%%%%%%%%%%%%%%%%%%%%%%%%%%%%%%%%%%%%%%%%%%%%%%%%%%%%%%%
\section{Timing analysis of the pulsar PSR~J1826$-$1334}
\label{radio}
The pulsar PSR~J1826$-$1334 (also known as PSR~B1823$-$13) was discovered in the survey
of ~\cite{clifton}. Its spin period of 101.48~ms, characteristic age of
21~kyr and spin-down power of $2.8\times 10^{36} \rm erg \, s^{-1}$ are
very similar to those of the Vela pulsar.
The distance to the pulsar as derived from the
dispersion measure is ($3.9 \pm 0.4$)~kpc~\citep{cor02}. 
Although the pulsar was not reported in the \emph{Fermi}-LAT catalog of 
gamma-ray pulsars with 6~months of data~\citep{psrcat}, we performed a temporal analysis 
on this increased \emph{Fermi}-LAT dataset using a new timing solution. 
A total of 162 observations of PSR J1826$-$1334 were made at 1.4 GHz 
using the Parkes \citep{wel09}, Lovell~\citep{jodrell} 
and Nan\c{c}ay \citep{nancay} radio telescopes.
The TEMPO2 timing package \citep{hob06} was then used to
build the timing solution. We fit the radio times of arrival (TOAs) to the pulsar rotation frequency 
and first three derivatives. We whitened the timing noise with four harmonically related sinusoids, 
using the ``FITWAVES'' functionality of the TEMPO2 package. The post-fit rms is
292.4~$\mu$s, or 0.2~\% of the pulsar phase. This timing solution will be made available 
through the Fermi Science Support Center~\footnote{FSSC: http://fermi.gsfc.nasa.gov/ssc/data/access/lat/ephems/}. 
 
Photons with an angle $\displaystyle \theta< \max(5.12^{\circ} \times 
(E/100 \, {\rm MeV})^{-0.8},0.2^{\circ})$, where $E$ is the energy of 
the photon, from the radio pulsar position, R.A. $=276.55490^{\circ}$, decl. $= -13.57967^{\circ}$ 
(J2000), were selected and phase-folded using the above mentioned ephemeris. The 
energy-dependence of the integration radius is a satisfactory 
approximation of the shape of the LAT Point Spread Function (PSF), especially at low energies.

The H-Test values, as defined in \citet{deJager1989}
and obtained from the analysis of the pulsed emission,
correspond to a significance well below $2 \sigma$ for each tested
energy band (30 MeV -- 300 GeV, 30 MeV -- 100 MeV, 100 MeV -- 300 MeV,
300 MeV -- 1 GeV, $>$ 1 GeV). No significant pulsation is detected
with the current statistics.  We fitted a point source at the position
of the pulsar PSR~J1826$-$1334 and derived an upper limit between
100~MeV and 1 GeV of $\sim 3.1 \times 10^{-8}$ cm$^{-2}$~s$^{-1}$, well
below typical gamma-ray fluxes reported for pulsars detected by
\emph{Fermi}-LAT~\citep{psrcat}. This ensures that any emission from the PWN HESS~J1825-137 
will not be contaminated by pulsed gamma-ray photons from PSR~J1826$-$1334.

%%%%%%%%%%%%%%%%%%%%%%%%%%%%%%%%%%%%%%%%%%%%%%%%%%%%%%%%%%%%%%%%%%%%%%%%%%%%%%%%%%%%%%%%%%%%%%%%%%%%%%%%%%%%%
\section{Analysis of the Pulsar Wind Nebula HESS~J1825$-$137}
\label{results}

The spatial and spectral analysis of the gamma-ray emission was
performed using two different methods, {\emph{gtlike}} and
{\emph{Sourcelike}}. {\emph{gtlike}} is the
maximum-likelihood method \citep{mat96} implemented in the
\emph{Fermi} SSC science tools. This tool fits a source model to the data along with models for the
instrumental, extragalactic and Galactic backgrounds. In the following
spectral analysis, the Galactic diffuse emission is modeled using the
ring-hybrid model {\it gll\_iem\_v02.fit}. The instrumental background
and the extragalactic radiation are described by a single isotropic
component with a spectral shape described by the tabulated model {\it
  isotropic\_iem\_v02.txt}.  The models and their detailed description
are released by the LAT Collaboration \footnote{$Fermi$ Science
 Support Center: http://fermi.gsfc.nasa.gov/ssc/}.  Sources within 10$^{\circ}$ of the pulsar PSR~J1826$-$1334 
 and found above the background with a statistical 
significance larger than $5 \, \sigma$ are extracted from the source
list given in \cite{FirstCat}, except for 1FGL~J1825.7$-$1410c which contributes partially
to the gamma-ray emission of the PWN. A detailed analysis reveals that the GeV emission is significantly
extended compared with that of a point source. The morphological
analysis was performed using \emph{Sourcelike}, which is described in the
Appendix. The extension test was done using a uniform disk and a
Gaussian spatial model. {\it Sourcelike} can also be used to assess
the Test Statistic (TS) value and to compute the spectra of both
extended and point-like sources. The TS is defined as twice the 
difference between the log-likelihood $L_1$ obtained by fitting a 
source model plus the background model to the data, and the log-likelihood 
$L_0$ obtained by fitting the background model only, i.e TS = 2($L_1$ $-$ $L_0$).

\subsection{Morphology}
\label{morpho}
\begin{figure*}[ht!!]
\begin{center}
\includegraphics[width=9.5cm]{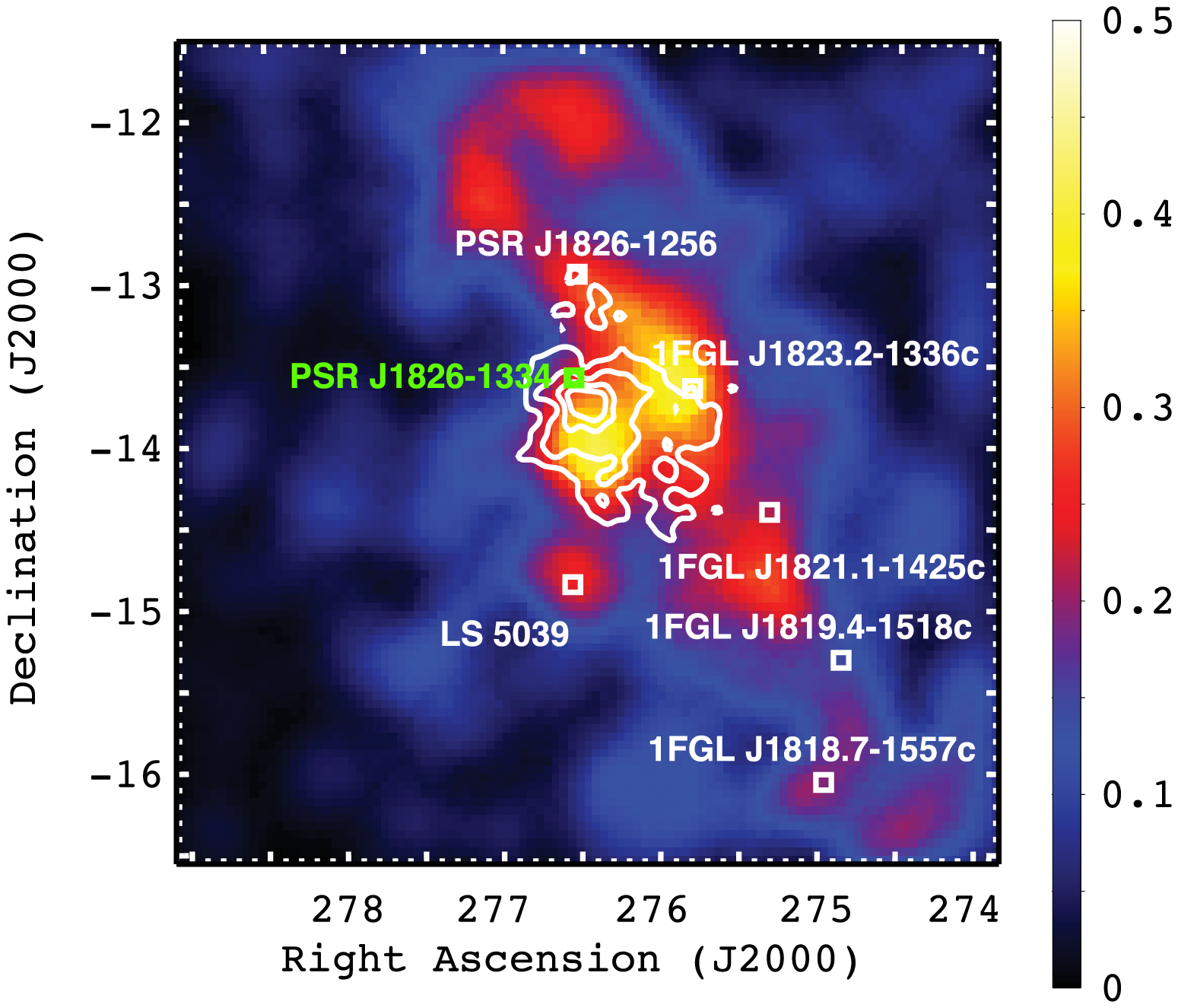}
\includegraphics[width=9.0cm]{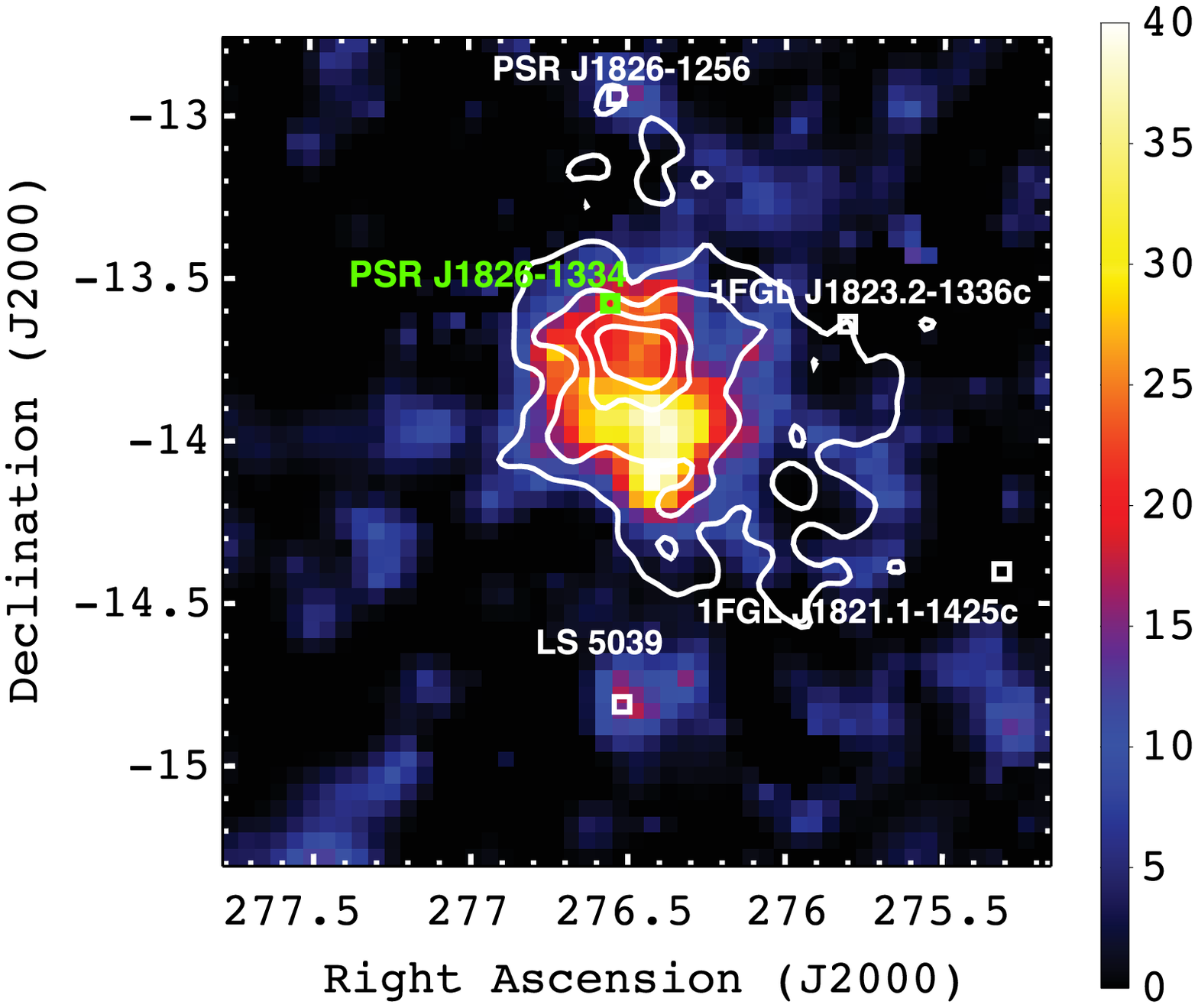}
\caption{Top: Fermi-LAT counts map above 10~GeV of the HESS~J1825$-$137 region with side-length
  $5^{\circ}$, binned in square pixels of side length 0.05$^{\circ}$. 
  The map is smoothed with a Gaussian of $\sigma =
  0.35^{\circ}$. H.E.S.S. contours \citep{HESSJ1825} are overlaid as
  gray solid lines. The position of the pulsar PSR~J1826$-$1334 and
  of the close-by 1FGL sources are indicated with green and white squares respectively. 
  LS~5039 is visible in the South-East at position (RA, Dec) =
  ($276.56^{\circ}$, $-14.85^{\circ}$). Bottom: Fermi-LAT Test Statistic (TS)
  map for events with energy larger than 10 GeV on a region of 2.5$^{\circ}$ side length. 
The TS was evaluated by placing a point-source at the center of each
pixel, Galactic diffuse emission and nearby sources being included in the background model.
}
\label{fig:maps}
\end{center}
\end{figure*}

To study the morphology of an extended source, a major
requirement is to have the best possible angular
resolution. Therefore, we decided to restrict our LAT dataset to
events with energies above 10~GeV. This also reduces the relative contribution of the Galactic
diffuse background.  Figure~\ref{fig:maps} (top) presents the LAT
counts map of gamma-ray emission around HESS~J1825$-$137, smoothed
with a Gaussian of $\sigma = 0.35^{\circ}$. The image contains emission from the Galactic
diffuse background and from nearby sources but in addition shows bright
emission south of PSR\, J1826$-$1334 coinciding generally with the
region that is bright at TeV energies (denoted by the
H.E.S.S. flux contours). The Test Statistc (TS, as defined above) map
above 10~GeV presented in Figure~\ref{fig:maps} (bottom) supports this
picture. This skymap contains the TS value for a point source at 
each map location, thus giving a measure of the statistical
significance for the detection 
of a gamma-ray source in excess of the background. Clearly, significant
emission to the south of PSR\,J1826$-$1334 is detected.

\begin{figure*}[ht!!]
\begin{center}
\includegraphics[scale=0.56]{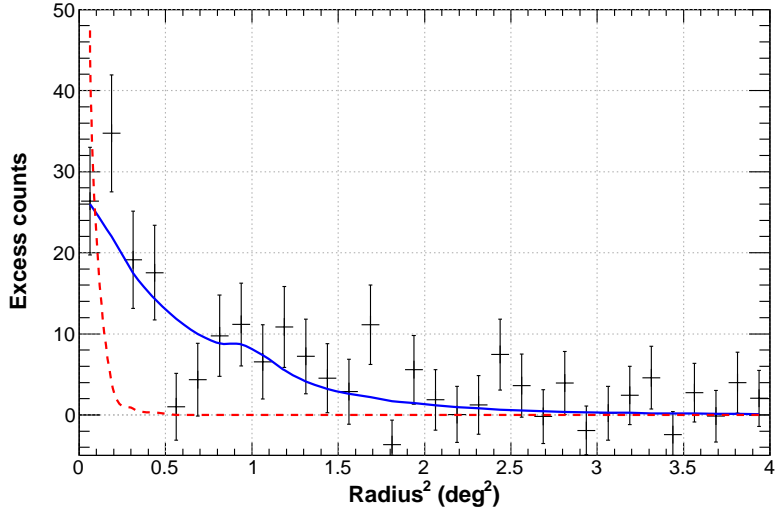}
\caption{ Background subtracted radial profile of the LAT data from the best-fit position provided by
Sourcelike for a Gaussian source (l, b) = (17$^{\circ}$.57, -0$^{\circ}$.43) as reported in Table~\ref{tab:sourcelike} 
(E $>$ 10 GeV). The best fit model, obtained for a Gaussian distribution, and the LAT PSF are overlaid as a blue 
solid and a red dashed lines respectively for comparison. The black dots represent the difference between the observed counts and the Galactic and extragalactic diffuse emission model. Nearby sources are not subtracted from the radial profile.
}
\label{fig:radial}
\end{center}
\end{figure*}

We determined the source extension using {\it Sourcelike} with a
uniform disk hypothesis and a Gaussian distribution (compared to the point-source hypothesis). The
results of the extension fits and the improvement of the TS
when using spatially extended models are summarized in
Table~\ref{tab:sourcelike}. The difference in TS between the Gaussian
distribution and the point-source hypothesis is $TS_{\rm ext} = 72$ (which converts
into a significance of $\sim 8 \sigma$ for the source extension) for
10~GeV $<$ $E$ $<$ 100~GeV, which demonstrates that the source is
significantly extended with respect to the LAT point spread function
(PSF). The fit extension has a dispersion of $\sigma = 0.56^{\circ} \pm 0.07^{\circ}$. 
We support this conclusion in Figure~\ref{fig:radial}, showing the background subtracted radial profile for the LAT data 
above 10 GeV (from the best source location determined for a Gaussian fit) and comparing this with the LAT PSF. 
Similar results are obtained assuming a uniform disk
model.

We have also examined the correlation of the gamma-ray emission with
different source morphologies by using {\it gtlike} with assumed
multi-frequency templates. For this exercise we compared the TS of the
point source and Gaussian distribution parameters provided by {\it
  Sourcelike} with values derived when using the H.E.S.S. gamma-ray excess map 
as a morphological template~\citep{HESSJ1825}.  The
resulting TS values obtained from our maximum likelihood
fitting are summarized in Table~\ref{tab:ts}. Fitting a uniform disk
to the data using the best location and size provided by {\it Sourcelike} instead of a point-source hypothesis results in $TS_{\rm ext} = 67$, comparable to the improvement in TS between
$D$ and $PS$ models in Table~\ref{tab:sourcelike}. Fitting a Gaussian
model improves the TS by 80. Replacing the Gaussian with spatial
template provided by the H.E.S.S. observations decreases the TS with
respect to the Gaussian hypothesis ($\Delta$TS = $-$42), implying
that the LAT emission is not perfectly reproduced by the H.E.S.S. excess map 
($E >$ 200 GeV). This is not completely surprising since H.E.S.S. 
reported an energy-dependent morphology for this source and \emph{Fermi} is probing 
lower energy electrons. Thus, while the best match is with the Gaussian morphology, we cannot rule out a simple disk morphology.

\placetable{1}
\begin{table}[ht]
\centering
\caption{\label{tab:sourcelike} Centroid and extension fits to the LAT
  data for HESS~J1825$-$137 using {\it Sourcelike} for events with
  energies above 10~GeV.  The difference in Test Statistic between a
  given spatial model and the point-source hypothesis is indicated by
  $TS_{\rm ext}$ in the last column.} 
\begin{tabular}{ccccccc}
\hline\hline 
Model & Name & l($^\circ$) & b ($^\circ$) & Radius ($^\circ$) & $TS_{\rm ext}$\\
\hline
Point Source & $PS$ & 17.62 & -0.82 & & \\
Disk & $D$ & 17.57 & -0.45 & 0.67 $\pm$ 0.02 & 72 \\
Gaussian & $G$ & 17.57 & -0.43 & 0.56 $\pm$ 0.07 & 72 \\
\hline
\end{tabular}
\end{table}

\begin{table}[ht]
\begin{center}
\caption{Comparison of model likelihood fitting results with {\it gtlike} for 
events with energies above 1~GeV. For each model, we give the name, 
and the Test Statistic value (TS).}\label{tab:ts}
\begin{tabular}{ccc}
\hline\hline
Model & Name & TS \\
\hline
Point Source & $PS$ & 24 \\
Disk & $D$ & 91 \\
Gauss & $G$ & 104 \\
H.E.S.S. & & 62 \\
\hline
\end{tabular}
\end{center}
\end{table}

\subsection{Spectral analysis}
\label{spectrum}
\begin{figure*}[ht!!]
\begin{center}
\includegraphics[angle=0,scale=.56]{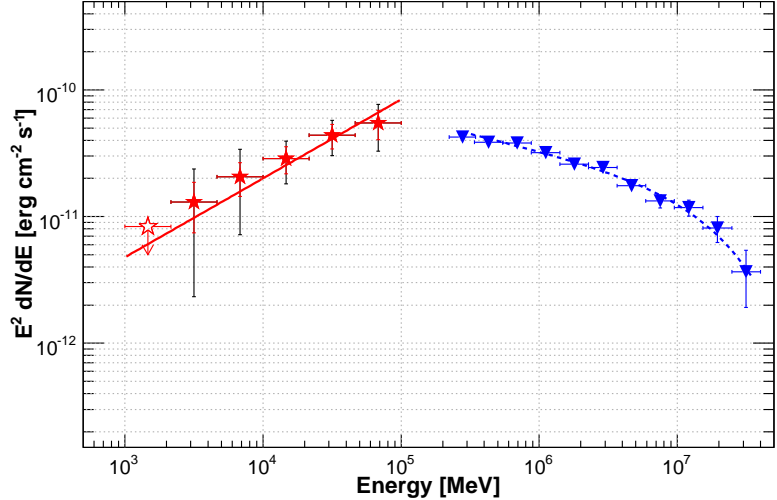} 
\caption{Spectral energy distribution of HESS~J1825$-$137 in gamma-rays.
  The LAT spectral points (in red) are obtained using the maximum likelihood
  method \emph{gtlike} described in section~\ref{spectrum} in 6
  logarithmically-spaced energy bins. The statistical errors are shown
  in red, while the black lines take into account both the statistical
  and systematic errors as discussed in section~\ref{spectrum}.  
The red solid line presents the result obtained by
  fitting a power-law to the data in the 1~--~100~GeV energy range
  using a maximum likelihood fit. A 95~\% C.L. upper limit is computed
  when the statistical significance is lower than 3~$\sigma$. The
  H.E.S.S. results are represented in blue \citep{HESSJ1825}.}
\label{fig:spec_hessJ1825}
\end{center}
\end{figure*}

The following spectral analyses are performed using \emph{gtlike}. 
The \emph{Fermi}-LAT spectral points were obtained by dividing the 1~--~100 GeV 
range into 6 logarithmically-spaced energy bins and
performing a maximum likelihood spectral analysis in each interval,
assuming a power-law shape for the source. For this analysis we used
the Gaussian model from Table~\ref{tab:sourcelike} to represent the
gamma-ray emission observed by the LAT, as discussed in
section~\ref{morpho}. Assuming this spatial shape, the gamma-ray
source observed by the LAT is detected with a TS of
104 ($\sim \, 10\sigma$) in the 1 -- 100 GeV range. To determine the integrated gamma-ray
flux we fit a power-law spectral model to the data in the energy range
1~GeV -- 100~GeV with a maximum likelihood analysis. This analysis is
more reliable than a direct fit to the spectral points since it accounts for Poisson
statistics of the data. The spectrum of HESS~J1825$-$137 between 1 and
100~GeV, assuming the Gaussian model from
Table~\ref{tab:sourcelike}, is presented in Figure~\ref{fig:spec_hessJ1825}. 
It is well described by a power-law with a spectral index of 1.38 $\pm$ 0.12 $\pm$ 0.16 
and an integral flux above 1~GeV of (6.50 $\pm$ 0.21 $\pm$ 3.90)$\times 10^{-9}$
cm$^{-2}$~s$^{-1}$. This is in agreement with results obtained
independently using {\it Sourcelike}. The first error is statistical,
while the second represents our estimate of systematic effects as
discussed below and is dominated by the uncertainties on the Galactic diffuse emission
in the 1~--~5~GeV energy range. With the current statistics, neither indication of a spectral cut-off at 
high energy nor significant emission below 1~GeV can be detected.

Four different systematic uncertainties can affect the LAT flux
estimation : uncertainties on the Galactic diffuse background, on the morphology of the LAT source, 
on the effective area and on the energy dispersion. The fourth one is relatively small 
($\le$~10\%) and has been neglected in this study. 
The main systematic at low energy is due to the
uncertainty in the Galactic diffuse emission since HESS~J1825$-$137 is
located only $0.7^{\circ}$ from the Galactic plane. Different versions
of the Galactic diffuse emission, 
generated by GALPROP~\citep{strong},  were used to estimate this
error. The observed gamma-ray intensity of nearby source free
regions on the galactic plane is compared with the intensity expected from the galactic
diffuse models. The difference, namely the local departure from the
best fit diffuse model, is found to be $\le 6$\%~\citep{w49}. 
By changing the normalization of the Galactic diffuse model artificially by $\pm
6$\%, we estimate the systematic error on the integrated flux of the PWN to be 70\% below 5 GeV, 
34\% between 5 and 10 GeV, and $<$12\% above 10~GeV. The second systematic
is related to the morphology of the LAT source. The fact that we do
not know the true gamma-ray morphology introduces another source of
error that becomes significant when the size of the source is larger than
the PSF, i.e above 600~MeV for the case of HESS~J1825$-$137. Different
spatial shapes have been used to estimate this systematic error: a
disk, a Gaussian distribution and the H.E.S.S. template. Our estimate of this
uncertainty is $\sim$30\% above 1~GeV.  The third uncertainty, common to every source analyzed with
the LAT data, is due to the uncertainties in the effective area.  This
systematic is estimated by using modified instrument response
functions (IRFs) whose effective area bracket that of our nominal
IRF. These `biased' IRFs are defined by envelopes above and below the
nominal dependence of the effective area with energy by linearly
connecting differences of (10\%, 5\%, 20\%) at log(E) of (2, 2.75, 4)
respectively.  We combine these various errors in quadrature to obtain
our best estimate of the total systematic error at each energy and
propagate through to the fit model parameters.

%%%%%%%%%%%%%%%%%%%%%%%%%%%%%%%%%%%%%%%%%%%%%%%%%%%%%%%%%%%%%%%%%%%%%%%%%%%%%%%
\section{Discussion}
\label{discussion}

A one-zone spectral energy distribution model provides a useful tool to investigate the
global properties of the PWN.
The nebula is non-uniform in the VHE regime
($\sim 1 ^{\circ} $ at 1 TeV and $\sim 0.2 ^{\circ}$ at 20 TeV)
and possesses a bright
central X-ray core observed with \emph{Chandra}~\citep{pavlov08}, \emph{XMM-Newton}
\citep{gaensler03} and \emph{Suzaku} \citep{uchi09} which extends no more than $15
\arcmin$ from the pulsar.
The non-uniform X-ray and VHE morphologies likely stem from cooling
losses by energetic electrons as they traverse the nebula, yet at
lower energies in the uncooled regime the
electron spectral shape remains essentially constant with time, and
hence also with position. As a result, even though a one-zone model
cannot reproduce the energy-dependent morphology of the nebula, a
one-zone model can nevertheless accurately reproduce the global flux
from uncooled electrons. Electrons inverse-Compton scattering off the
CMB require energies of $\sim 2$ TeV in order to produce photons of
mean energy 10 GeV in the midst of the LAT energy range.  Even in a
$10 \, \mu \rm{G}$ magnetic field such electrons synchrotron cool
quite slowly over a timescale of $\sim 45$ kyr, roughly double the
characteristic age of the pulsar.  Inspection of Figure~\ref{fig:spec_hessJ1825} indicates a
spectral break at $\sim 200 \, \rm{GeV}$, almost certainly due to a
cooling break in the electron spectrum.  The hard LAT spectrum is
therefore clearly in the uncooled regime, and so a one-zone model can
help illuminate this new data.

We apply a one-zone time dependent spectral energy distribution (SED) model, as described in
\cite{velax}. This model computes SEDs from evolving
electron populations over the lifetime of the pulsar in a series of
time steps, with the energy content of the injected particle
population varying with time following the pulsar spin down. 
During the free-expansion phase of the PWN (assumed to be
$\sim 10^4$ years) we adopt an expansion of $R \propto t$,
following which the radius $R \propto t^{0.3}$, appropriate
for a PWN expanding in pressure equilibrium with a Sedov phase
SNR.  
Over the pulsar lifetime the magnetic field
$B \propto t^{-0.5}$, following $\sim$ 500 years of constancy.  
At each time step synchrotron, inverse-Compton (Klein-Nishina effects included),
and adiabatic losses are calculated. Synchrotron and IC fluxes are 
calculated from the final electron spectrum. We allow the
braking index $n$ of the pulsar to vary, thereby changing the age and
spin-down behavior of the pulsar.

We assume the existence of three primary photon fields 
(CMBR, far IR (dust), and starlight) and   
use the interstellar radiation mapcube within the GALPROP suite \citep{porteretal05} 
to estimate the photon fields at the Galactic radius of PSR\,
J1826$-$1334. A distance of 3.9~kpc in the direction of the pulsar corresponds to 
a Galactic radius of 4.7~kpc. At this radius, the peak of the SED of dust IR photons corresponds to a black body
temperature of $T \sim 32$ K with a density of $\sim 0.9$
eV$\rm \, cm^{-3}$, while the SED of stellar photons peaks at $T \sim 2500$ K
with a density of $\sim 3.6$ eV$\rm \, cm^{-3}$.

Spectral measurements consist of LAT and H.E.S.S.\ data points, as well as an estimate
of the X-ray spectrum. We adopt an X-ray photon index of $2.2 \pm 0.3$ and a flux 
of $(7 \pm 2) \times 10^{-12} \, {\rm erg \, cm^{-2} \, s^{-1}}$.  The selected index
is consistent with the indices measured by \emph{XMM-Newton} \citep{gaensler03}
and \emph{Suzaku} \citep{uchi09} for the extended nebula.  The flux level 
is equivalent to sum of all the \emph{Suzaku} regions analyzed by
\citet{uchi09}, with 30 \%
systematic errors assumed. 

\begin{figure*}[ht!!]
\begin{center}
\includegraphics[angle=0,scale=0.6]{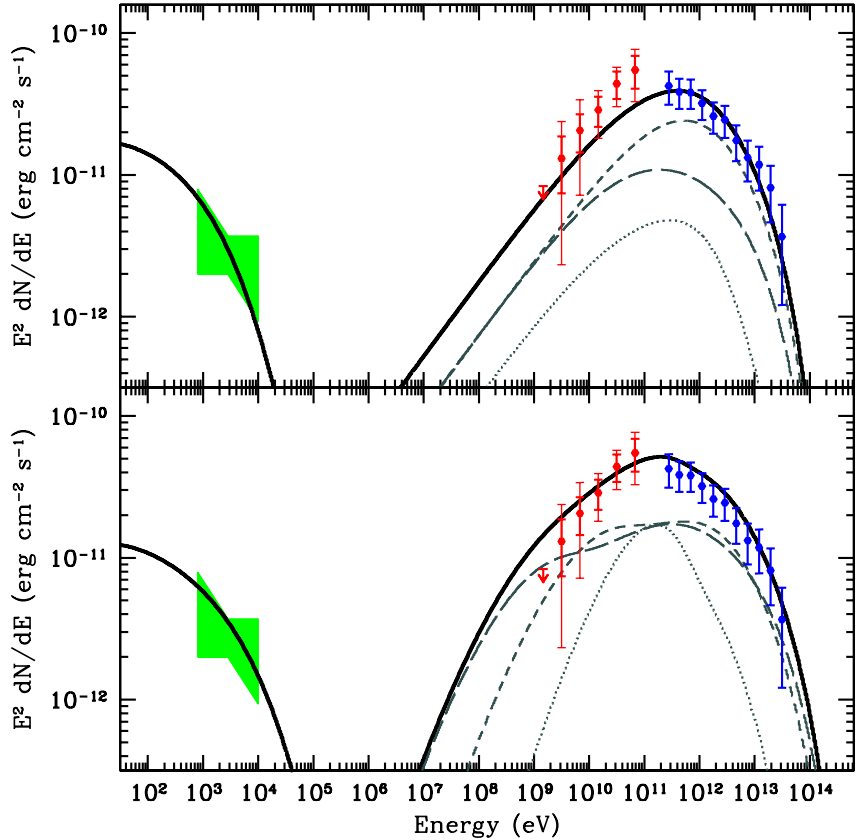}
\caption{\label{fig:powmodel}Spectral energy distribution of
HESS~J1825$-$137 with a simple exponentially cutoff power-law electron spectrum (Top), 
and a relativistic Maxwellian plus 
power-law electron spectrum (Bottom). The LAT spectral points (red, thin lines
denote systematic errors) H.E.S.S. points 
(blue), and X-ray bowtie (green) are shown. 
The black line denotes the total synchrotron and inverse Compton emission from the nebula.
Thin curves indicate the Compton components from scattering on the CMB (long-dashed), 
IR (medium-dashed), and stellar (dotted) photons.}
\end{center}
\end{figure*}

A simple exponentially cutoff power-law injection of electrons,
evolved properly over the pulsar lifetime, often provides an adequate
match to PWNe SEDS. Initially, we fit this injection
spectrum with four variables: final magnetic field, electron high
energy cutoff, electron power-law index, and the pulsar braking index
$n$.  Given the large covariance between the braking index and the
initial spin period in determining the age of the pulsar, we fix the
initial spin period at 10 ms and braking index at 2.5, yielding an age 
of 26 kyr for the system. This
simple injection spectrum slightly underestimates the LAT data but the overall fit is still 
reasonable. For the source age of 26~kyr, we require a power-law index of $1.9$, a cutoff 
at $57$~TeV and a magnetic field of $4 \, \mu \rm{G}$. 
The corresponding result is presented in Figure~\ref{fig:powmodel} (Top).

Another option to fit the multi-wavelength data is adopting the relativistic Maxwellian plus power-law
tail electron spectrum proposed by \citet{spitkovsky08}. For this
injection spectrum, we assume a bulk gamma-factor ($\gamma_0$) for the
PWN wind upstream of the termination shock.  At the termination shock
the ambient pressure balances the wind pressure, fully thermalizing
the wind; in this case the downstream post-shock flow has $\gamma =
(\gamma_0 - 1)/2$. 
One could also interpret this as an effective temperature $kT$ of $m_e c^2 \,
(\gamma_0 - 1) /2$. Per the simulations of \citet{spitkovsky08}, a
power-law tail begins at $7 kT \sim 7/2 m_e c^2 \, \gamma_0$, and
suffers an exponential cutoff at some higher energy.  For our modeling
we fix the power-law beginning at $\sim 7 kT$, and allow $kT$, the
power-law index, and the exponential cutoff to vary. The best fit, presented in Figure~\ref{fig:powmodel} (Bottom), 
is obtained with $kT = 0.14$ TeV, corresponding to an upstream gamma-factor of 
$5.5 \times 10^{5}$, a magnetic field of $3 \, \mu \rm{G}$, a cutoff at $150$~TeV and a power-law index of $2.3$ 
close to the value of $\sim 2.5$ proposed by \citet{spitkovsky08}. The relativistic Maxwellian plus power law
model matches the multi-wavelength data and also directly probes the upstream 
pulsar wind via fitting of $\gamma_0$.

HESS~J1825$-$137 is detected at high significance by the Fermi LAT, and 
demonstrates both morphological similarity and flux continuity with 
the H.E.S.S. regime. The LAT spectral index of $1.38 \pm 0.12 
\pm 0.16$ is consistent with both a simple power-law electron
injection spectrum, as well as a Maxwellian plus power-law injection spectrum over a 
simple power-law. A mean magnetic field of $\sim 3 - 4 \, \mu$G adequately
fits the X-ray flux, and an age of $\sim 26$ kyr is consistent with the data.
A total of $5 \times 10^{49} \, \rm{erg}$ injected in the form of electrons by the pulsar
is required to match the gamma-ray flux in the nebula.  

%%%%%%%%%%%%%%%%%%%%%%%%%%%%%%%%%%%%%%%%%%%%%%%%%%%%%%%%%%%%%%%%%%%%%%%%%%%%%%%%%%%%%%%%%%%%%%%%%%%%%%%%%%%%%%%%%%%%%%%%%%%%%%%%%%%
\footnotesize
\acknowledgments
The \emph{Fermi} LAT Collaboration acknowledges generous ongoing
support from a number of agencies and institutes that have
supported both the development and the operation of the LAT as well as 
scientific data analysis. These include the
National Aeronautics and Space Administration and the Department
of Energy in the United States, the Commissariat \`a
l'Energie Atomique and the Centre National de la Recherche
Scientifique / Institut National de Physique Nucl\'eaire et de
Physique des Particules in France, the Agenzia Spaziale Italiana,
the Istituto Nazionale di Fisica Nucleare, and the Istituto
Nazionale di Astrofisica in Italy, the Ministry of Education,
Culture, Sports, Science and Technology (MEXT), High
Energy Accelerator Research Organization (KEK) and Japan
Aerospace Exploration Agency (JAXA) in Japan, and the K.
A. Wallenberg Foundation and the Swedish National Space
Board in Sweden.
Additional support for science analysis during the operations 
phase from the following agencies is also gratefully acknowledged: 
the Instituto Nazionale di Astrofisica in Italy and the Centre National d'\'Etudes Spatiales in France.\\
The Nan\c{c}ay Radio Observatory is operated by the Paris Observatory, associated with the French Centre National de la Recherche Scientifique (CNRS).\\
The Lovell Telescope is owned and operated by the University of Manchester as part of the Jodrell Bank Centre for Astrophysics with support from the Science and Technology Facilities Council of the United Kingdom.\\
The Parkes radio telescope is part of the Australia Telescope which is funded by the Commonwealth Government for operation as a National Facility managed by CSIRO. We thank our colleagues for their assistance with the radio timing observations.

\normalsize
\section*{Appendix : Description of \emph{Sourcelike}}

\emph{Sourcelike} is a tool developed for performing morphological studies
of spatially extended Fermi sources. \emph{Sourcelike} is an extension
to \emph{pointfit} \citep[described in][]{FirstCat}, which was developed to efficiently create Test
Statistic maps and localize catalog sources with little sacrifice
of precision.  \emph{pointfit} bins the sky in position and energy
and increases efficiency by scaling the spatial bin size with energy.
Furthermore, it uses a region of the sky centered on the source
whose radius is energy dependent: from $15^{\circ}$ at 100~MeV to
$3.5^{\circ}$ at 50~GeV.  It was successfully used by and is described
in the 1FGL catalog \citep{FirstCat}.  \emph{Sourcelike} deviates from
\emph{pointfit} by fitting not the PSF but the PSF convolved with an
assumed spatial shape. By independently fitting the flux in each
energy bin, \emph{Sourcelike} performs an extension analysis without
biasing the fit by assuming a spectral model. Since the PSF ranges
a full two orders of magnitude in size over the energy range of the
instrument, this maximum likelihood approach naturally handles the
energy dependent PSF and maximizes our sensitivity to extension.

To find the shape of the source that most closely matches the
observed photons, the overall likelihood is maximized by simultaneously
fitting the spatial parameters of the source. At each step in the
fit, the new shape must be again convolved with the PSF. It is for
this reason that this optimized software was developed. Errors on
the fit position and extension are estimated by numerically calculating
the curvature of the likelihood function at the best fit spatial
model.  The statistical significance of a source's extension is
computed by \emph{Sourcelike} by calculating $\rm{TS}_{ext}$,
which is defined as twice the difference between the log-likelihood
$L_1$ obtained by fitting as an extended source and the log-likelihood
$L_0$ obtained by fitting as a point source, i.e $\rm{TS} =
2(\rm{L}_1- \rm{L}_0$). The statistical significance of a source's
extension is then calculated as $\sqrt{\rm{TS}_{ext}}$.
\emph{Sourcelike} is particularly well suited for extension studies. It was
extensively tested and validated against
{\it gtlike}, and has been used in other LAT studies of PWNe \citep{velax,msh1552}.

%% The following command ends your manuscript. LaTeX will ignore any text
%% End of file 

\end{document}